\documentclass[11pt,a4paper]{emulateapj}
\usepackage{natbib}
\usepackage{amssymb}
\usepackage{amsmath} 
\usepackage{amsbsy}
\usepackage{slashbox}
\usepackage{color}

\usepackage{epsfig}
\usepackage{amsmath}
\usepackage{amssymb}
\usepackage{natbib}
\usepackage{subfigure}
\usepackage{graphicx}

\slugcomment{submited to {\it The Astrophysical Journal}}

\newcommand{\civ}{C {\sc iv}}

\newcommand{\heii}{He {\sc ii}}
\newcommand{\ciii}{C {\sc iii}]}

\begin{document}

\def\mean#1{\left< #1 \right>}

\title{Discovery of an Enormous Ly$\alpha$ nebula in a massive galaxy overdensity at \lowercase{$z}=2.3$}
\author{Zheng Cai\altaffilmark{1,2,11}, Xiaohui Fan\altaffilmark{2}, Yujin Yang\altaffilmark{3},  Fuyan Bian\altaffilmark{4},  J. Xavier Prochaska\altaffilmark{1}, Ann Zabludoff\altaffilmark{2}, 
  Ian McGreer\altaffilmark{2}, Zhen-Ya Zheng\altaffilmark{5,6}, Richard Green\altaffilmark{2},   Sebastiano Cantalupo\altaffilmark{7}, Brenda Frye\altaffilmark{2},  Erika Hamden\altaffilmark{8}, Linhua Jiang\altaffilmark{9},  Nobunari Kashikawa\altaffilmark{10}, Ran Wang\altaffilmark{9}}
\affil{$^1$ UCO/Lick Observatory, University of California, 1156 High Street, Santa Cruz, CA 95064, USA }
\affil{$^2$ Steward Observatory, University of Arizona, 933 North Cherry Avenue, Tucson, AZ, 85721, USA}
\affil{$^3$ Korea Astronomy and Space Science Institute, 776 Daedeokdae-ro, Yuseong-gu Daejeon, Korea}
\affil{$^4$ Research School of Astronomy \& Astrophysics, Mount Stromlo Observatory, Cotter Road, Weston ACT 2611, Australia}
\affil{$^5$ Instituto de Astrofisica, Pontificia Universidad Catolica de Chile, 7820436 Santiago, Chile}
\affil{$^6$ Chinese Academy of Sciences South America Center for Astronomy, 7591245 Santiago, Chile}
\affil{$^7$ Institute for Astronomy, ETH Zurich, Wolfgang-Pauli-Strasse 27, 8093 Zurich, Switzerland}
\affil{$^8$ California Institute Technology, 1200 East California Boulevard, Pasadena, CA, USA}
\affil{$^9$ The Kavli Institute for Astronomy and Astrophysics, Peking University, Beijing 100871, P. R. China}
\affil{$^{10}$ National Astronomical Observatory of Japan, Mitaka, Tokyo,181-8588, Japan}
\affil{$^{11}$ Hubble Fellow}


\altaffiltext{1} {Email: zcai@ucolick.org}

\begin{abstract}
Enormous Ly$\alpha$ Nebulae (ELANe), unique tracers of galaxy density peaks, are predicted to lie at the nodes and intersections of cosmic filamentary structures. Previous successful searches for ELANe have focused on wide-field narrowband surveys, or have targeted known sources such as ultraluminous quasi-stellar-objects (QSOs) or radio galaxies. 
Utilizing groups of coherently strong Ly$\alpha$ absorptions (CoSLAs), we have developed a new method to identify high-redshift galaxy overdensities and have identified an extremely massive overdensity, BOSS1441, at $z=2-3$  (Cai et al. 2016a). In its density peak, we discover an ELAN that is associated with a relatively faint continuum. To date, this object has the highest diffuse Ly$\alpha$ nebular luminosity of $L_{\rm{nebula}}=5.1\pm0.1\times10^{44}$ erg s$^{-1}$.  Above the 2$\sigma$ surface brightness limit of SB$_{\rm{Ly\alpha}}= 4.8\times10^{-18}$ erg s$^{-1}$ cm$^{-2}$ arcsec$^{-2}$, this nebula has an end-to-end spatial extent of 442 kpc. This radio-quiet source also has extended \civ\ $\lambda1549$ and \heii\ $\lambda1640$ emission on $\gtrsim30$ kpc scales. 
Note that the Ly$\alpha$, \heii\ and \civ\ emission all have double-peaked line profiles. Each velocity component has a full-width-half-maximum (FWHM) of $\approx700 - 1000$ km s$^{-1}$. We argue that this Ly$\alpha$ nebula could be powered by shocks due to an AGN-driven outflow or/and photoionization by a strongly obscured source. 

\end{abstract}

\section{Introduction}

During the peak epoch of galaxy formation at $z=2-3$ \citep[e.g.,][]{bouwens11}, most of the baryons in the Universe reside outside galaxies; they lie within the intergalactic medium (IGM) and circumgalactic medium (CGM) \cite[e.g.,][]{cen94, miralda96, hernquist96, rauch98}. The IGM and CGM provide a vast reservoir for fueling the star formation of galaxies and also serve as a ``sink" for metals driven from the galactic feedback \cite[e.g.,][]{prochaska11, tumlinson11}. On the other hand, the properties and structures of the IGM/CGM, such as kinematics, morphology, and metallicity, are increasingly reshaped by the energetic processes occuring in galaxies, and therefore the IGM/CGM acts as a laboratory to stringently constrain the physics of the galaxy formation \cite[e.g.,][]{fumagalli11}.  

Intergalactic/circumgalactic filaments have been studied via QSO absorption lines \cite[e.g.,][]{rauch98}. But QSO absorption studies are limited due to the sparseness of background QSOs. 
To reveal the connection of intergalactic gas to galaxies and their circumgalactic medium (i.e. on scales of $\sim$100 kpc), one must constrain the  full three-dimensional intergalactic/circumgalactic material using more numerous, but fainter, background galaxy populations \cite[e.g.,][]{lee14} or directly map the faint diffuse emission of the intergalactic medium (IGM) or circumgalactic medium (CGM) \cite[e.g.,][]{cantalupo14, martin15, borisova16}. The Ly$\alpha$ line is the primary coolant of gas with $T\sim 10^4$ K and can be used to trace the CGM/IGM via emission. Such Ly$\alpha$ nebulae provide us an indispensable opportunity to study the CGM in emission.

Theoretical models suggest that several mechanisms may generate circumgalactic Ly$\alpha$ emission: (1) recombination radiation following photoionization (fluorescence) powered by ultraviolet (UV) sources \citep{gould96, cantalupo05, geach09, kollmeier10}; (2) cooling radiation due to the gravitationally heated gas \citep{fardal01, yang06, dijkstra09, faucher10}; (3) radiation from shock-heated gas driven by the feedback of galactic outflow \cite[e.g.,][]{villar07, taniguchi00, wilman05}; and (4) resonant scattering of Ly$\alpha$ from the embedded source \citep{dijkstra09, hayes11, cantalupo14, geach14, geach16}.  The photoionization radiation is generated when the dense regions of the CGM are photoionized by strong ionizing sources and then recombine to emit Ly$\alpha$ photons. Cooling radiation is the Ly$\alpha$ photons released when gas settles into galactic potential wells \cite[e.g.,][]{yang06}. Shock-heating can be powered by supernovae, or by relativistic winds or jets resulting from gas accretion onto supermassive black holes (SMBHs). 
 Ly$\alpha$ resonant scattering produces extended Ly$\alpha$ halos as Ly$\alpha$ photons propagate outward and is characterized by a double-peaked structure of the resonant emission lines \cite[e.g.,][]{yang14}. 
These mechanisms are believed to power the extended Ly$\alpha$ emission in high-density regions of the early Universe.  The Ly$\alpha$ nebulae/blobs (LABs) are expected to occupy massive dark matter halos ($\sim$ 10$^{13}$ M$_\odot$), representing sites of the most active star formation and tracing large-scale mass overdensities \cite[e.g.,][]{steidel00, prescott09, yang09}.

A few observational efforts have been made to search for Ly$\alpha$ nebulae/blobs at $z=2-3$.  These successful searches   include narrowband imaging surveys of random fields \cite[e.g.,][]{steidel00, francis01, palunas04, dey05, yang09, prescott09, yang10}, narrowband imaging of known overdensities \citep{matsuda05}, and targeting biased halo tracers, such as ultraluminous QSOs \citep[e.g.,][]{cantalupo14, hennawi15} and radio galaxies \cite[e.g.,][]{heckman91, villar07, miley08}. 
Using VLT/MUSE, \citet{borisova16} present a blind survey for Ly$\alpha$ nebulae associated with 17 brightest radio-quiet QSOs at $3 < z < 4$. They find that 100\% of the QSOs are associated with Ly$\alpha$ nebulae with linear sizes of $\sim100$ -- 320 kpc. In this sample, the relatively narrow Ly$\alpha$ FWHMs (300 -- 600 km s$^{-1}$)  are consistent with a fluorescent powering mechanism. 
Increasing evidence has shown that the Ly$\alpha$ nebulae often lie in regions that contain both  enhanced UV-radiation (or nearby UV sources) and gas overdensities \citep{hennawi13, hennawi15}.

The extended \heii\ and \civ\ associated with Ly$\alpha$ nebulae contain crucial information about the powering mechnisms. The extended \civ\ line allows us to estimate the metallicity of the CGM gas and the size of the metal enriched halos \citep{arrigoni15a}. In turn, such metal line emission allows us to examine whether the shocks of the galactic outflow could power the LABs \cite[e.g.,][]{villar99, villar07, allen08, arrigoni15b}. 
Arrigoni-Battaia et al. (2015a) conducted a deep survey of 13 Ly$\alpha$ blobs in the SSA22 overdensity (Steidel et al. 2000; Matsuda et al. 2005), targeting the \heii\ $\lambda1640$ and \civ\ $\lambda1549$. These observations did not detect extended \heii\ and \civ\ emission in any of the LABs, suggesting that photoionization could be a major powering mechanism. \citet{borisova16} also did not detect strongly extended He II and C IV emission in their sample of the 17 ultraluminous QSOs, indicating a large fraction of the gas in massive QSO host halos at $z=3-4$ could be cold ($T\sim10^4$ K) and metal-poor ($Z < 0.1 Z_\odot$). 
\citet{prescott09} detect a LAB that has a spatial extent of 80 kpc at $z\approx1.67$ associated with extended \civ\ and \heii. The Ly$\alpha$, \civ, \heii\ and \ciii lines all show a coherent velocity gradient of 500 km s$^{-1}$, strongly indicating a 50 kpc large rotational disk illuminated by an AGN. 

Recently, two enormous Ly$\alpha$ nebulae (ELANe) have been discovered to have a large spatial extent of $\gtrsim400$ kpc \citep{cantalupo14, hennawi15}.  
 These ELANe further offer excellent laboratories to detect and map the gas in the dense part of the intergalactic medium (IGM), and to study how the IGM feeds star formation in massive halos \citep{martin14, martin15}. \citet{arrigoni15a} conducted deep spectroscopic integrations targeting \heii\ and \civ\ emission and report a null detection, suggesting ELANe are mainly due to AGN photoionization on the cool,  metal-poor CGM gas. 

In this paper, we report a discovery of another ultraluminous ELAN that resides near the density peak of our newly discovered massive overdensity BOSS1441 at $z=2.32\pm0.02$ (Cai et al. 2016a). This nebula has a projected linear size of $\approx 450$ kpc, comparable with the Slug nebula \citep{cantalupo14}, and remarkably extended  \heii\ and \civ\ emission over $\gtrsim30$ kpc. The Ly$\alpha$, \heii\ and \civ\ lines all show double-peaked kinematics, with each component having the line widths of 700 $-$ 1000 km s$^{-1}$.
The large spatial extent of  Ly$\alpha$ emission, the strongly extended \heii\ and \civ, and the emission line structures and kinematics all make this ELAN unique. This Ly$\alpha$ nebula resides in an overdense field selected utilizing the  largest QSO spectral library from the  Baryon Oscillations Spectroscopic Survey (BOSS) \cite[e.g.,][]{dawson13}. It contains a group of extremely rare, high optical depth Ly$\alpha$ absorption (Cai et al. 2015) arising from the IGM overdensity and a rare QSO group \cite[e.g.,][]{cai16a}. We refer to this program as MApping the Most Massive Overdensity Through Hydrogen (MAMMOTH) (Cai et al. 2015).  In this paper, we refer this nebula as MAMMOTH-1.

 This paper is structured as follows. In \S2, we introduce the selection of MAMMOTH-1 and our follow-up observations.  In \S3, we discuss our observational results.  In \S4, we discuss the physical properties and several powering mechanisms that could be responsible for such a unique ELAN. We also estimate the cool gas mass. We give a brief summary in \S 5.  We convert redshifts to physical distances assuming a $\rm{\Lambda}$CDM cosmology with $\Omega_m= 0.3$, $\Omega_{\Lambda}=0.7$ and $h=0.70$ ($h_{70}$).  Throughout this paper when measuring distances, we normally refer to physical separations or distances. We use cMpc to represent comoving Mpc, and kpc to represent physical kpc.

\section{Observations}

\subsection{Target Selection}

 MAMMOTH-1 is located in the density peak of the large-scale structure BOSS1441 (Cai et al. 2016a). BOSS1441 was selected because this field contains a group of 5 strong Ly$\alpha$ absorption systems within a 20 $h^{-1}$ comoving Mpc (cMpc) scale at $z=2.32\pm0.03$. Each Ly$\alpha$ absorption has an effective optical depth on a scale of 15 $h^{-1}$ cMpc of  $\tau_{\rm{eff}}^{15h^{-1}{\rm{Mpc}}}>3\times$ the mean optical depth ($\mean{\tau_{\rm{eff}}}$). These absorption systems are not due to DLAs. Two of them have $\tau_{\rm{eff}}^{15h^{-1}{\rm{Mpc}}}>4.5\times \mean{\tau_{\rm{eff}}}$ and the optical depth is higher than the threshold of coherently strong Ly$\alpha$ absorption (CoSLA, see Cai et al. 2016a). This group of absorbers satisfies the selection criteria (a) -- (d2) proposed in Cai et al. (2015). 
 The redshift is chosen by our custom narrowband filter {\it NB403}. 

The NB403 filter has a central wavelength of $\lambda_c = 4030$ \AA\ and a bandwidth of FWHM $=45 $ \AA. The NB filter is very efficient to search for the overdensities, because (1) the BOSS QSO density peak lies at $z\sim2.3$. With a NB filter at a similar redshift, we can fully take advantage of the SDSS Ly$\alpha$ forest survey; (2) The KPNO-4m/MOSAIC camera is highly sensitive at $\sim$ 4000 \AA. In addition, $z\sim2.3$ is a good redshift for studying galaxy properties using ground-based telescopes. Optical and infrared spectrographs can fully cover the emission lines from Ly$\alpha$ to H$\alpha$.

\subsection{KPNO-4m/MOSAIC Narrowband + Broadband Imaging}

After selecting BOSS1441 field, we conducted deep narrowband + broadband imaging and multi-object spectroscopy to select and confirm Ly$\alpha$ emitting galaxies (LAEs). 
We used the KPNO-4m/MOSAIC-1.1 camera for deep imaging with a custom narrowband filter {\it NB403} and the Bw broadband filter. These deep imaging observations were designed to reveal LAEs in the BOSS1441 field. 

We  briefly review our observations in BOSS1441 field. More details can be found in Cai et al. (2016a). 
The BOSS1441 field  was observed on Mar.~2013, Apr.~2014, and Jun.~2014. For the NB403 filter, the total exposure time was 17.9 hours, which consisted of individual 15 or 20 minute exposures with a standard dither pattern to fill in the
gaps between the eight MOSAIC CCD chips. The seeing ranged from $1.1''- 1.7''$, with the median seeing about $1.32''$. For the Bw filter, the total exposure time is 3 hours, which consists of individual 8 minute exposure with fill gap dither pattern. The seeing for taking Bw band ranges from $1.1'' - 1.8''$, with the median value of 1.37$''$. 
Around the LAE density peak region, we discovered strong, highly extended Ly$\alpha$ emission: the ELAN MAMMOTH-1, which is detected in each individual frame with a 15 -- 20 min exposure. 
The broadband filter (Bw) was observed for total 3 hours to match the depth of narrowband filter.  These observing conditions enable us to achieve a narrowband magnitude $m_{\rm{NB403}}=25.1$ at 5$\sigma$ (aperture: 2.5''), and Bw-band magnitude $m_{\rm{Bw}}=25.9$ at 5$\sigma$. This depth corresponds to a $1\sigma$ Ly$\alpha$ surface brightness limit of SB$_{\rm{Ly\alpha}}= 2.4\times10^{-18}$ erg s$^{-1}$ cm$^{-2}$ arcsec$^{-2}$.

\subsection{LBT Imaging and Multiobject Spectroscopy}

We used the Large Binocular Camera (LBC) on the Large Binocular Telescope (LBT) to obtain deep imaging using the $U$, $V$, and $i$ broadband filters. The LBC imaging enables the selection of the star-forming galaxies at $z\approx2.3$ \cite[e.g.,][]{adelberger05} and helps to eliminate [OII] contaminants using the BX galaxy selection technique. 

We also used the Dual channel of the Medium-Dispersion Grating Spectroscopy (MODS) \citep{byard00} on the LBT to  spectroscopically confirm the redshifts of galaxies and the Ly$\alpha$ nebula in the BOSS1441 overdense field. The LBT/MODS provides high efficiency over 3200\AA\ -- 10,000\ \AA\ with a resolution of $R= 2000$. 
We used a dichroic that divides the incoming beam at $\approx$ 5700\AA. This configuration covers Ly$\alpha$ and a few interstellar lines, e.g., \civ\ $\lambda1550$, \heii\ $\lambda1640$, \ciii\ $\lambda1907/1909$, for galaxies at $z=2.3$. 

We used a MODS mask to observe the MAMMOTH-1 nebula and other LAEs in the field. The total exposure time was 6 hours. Each mask was split into six 1,800 sec integrations, with a typical seeing of 1$''$. The slit length on the MAMMOTH-1 nebula is $10''$ (Figure~1). 

The MODS data reduction followed the LBT/MODS reduction routine.  First, each raw image was processed with the MODS CCD  reduction utilities (modsTools v03) to obtain bias-subtracted and
 flat-fielded images. We generated polynomial
fits to the arc calibration to determine the transformation between image
pixels and wavelength. The sky model was fit to each image using B-splines and then subtracted. 
We used LACOSMIC (Van Dokkum 2010) to identify cosmic rays during the construction of the sky model. The
individual exposures were combined with inverse variance
weighting to produce the final 2D spectrum. 

{\section{Observational Results}}

{\subsection{Mapping the Ly$\alpha$ Emission}}

In Figure~1, we present the stacked images of MAMMOTH-1 in both the NB403 narrowband and Bw broadband images.
We also overplot the LBT/MODS spectral slit. From this figure, we detect extended structures in both the narrowband (NB403, left panel) and broadband images (Bw, right panel). 
In the Bw broadband, we detect multiple sources associated with the MAMMOTH-1.
In Figure 2, we present the continuum subtracted Ly$\alpha$ image. We smooth the image using a Gaussian Kernel with $1''$ \citep{cantalupo14, hennawi15}.  
 Within the 2$\sigma$ ($4.8\times10^{-18}$ erg s$^{-1}$ cm$^{-2}$ arcsec$^{-2}$) surface brightness contour, this ELAN has an end-to-end projected extent of 53 arcsec (442 physical kpc).

 In the broadband image, the brightest two sources: brighter source A ($B_{\rm{AB}}\approx 23.5,\  i_{\rm{AB}}\approx22.5$) and fainter source B ($B_{\rm{AB}}\approx 25.1$, $i_{\rm{AB}}\approx24.3$) are marked in Figure 1. Source B resides in the flux peak of the broad-band subtracted narrowband image. Our LBT/MODS spectroscopy shows that source A is a low-redshift AGN at 
$z=0.16$, while source B is an object at $z=2.32$. In Figure~3, we present the 1-D spectrum of source B which has strong emission in Ly$\alpha$, \heii, \civ, and \ciii. Using LBT/LBC imaging, we find that source B has a brightness of $U_{\rm{AB}}=25.77\pm0.07$,  $V_{\rm{AB}}=24.37\pm0.03$, $i_{\rm{AB}}=24.30\pm0.03$. Although it is difficult to identify all the possible powering sources  associated with MAMMOTH-1, source B's location and redshift suggest that it could be the dominant powering source of MAMMOTH-1. 
We use source B's position as the center of the MAMMOTH-1: $\alpha$ = 14:41:24.475 $\delta=$+40:03:09.45.

From the broadband-subtracted narrowband image (Figure 2), we measure that MAMMOTH-1 has a total Ly$\alpha$ luminosity of $5.28\pm0.07\times10^{44}$ erg s$^{-1}$.
Unlike ELANe powered by ultraluminous type-I QSOs, 
the Ly$\alpha$ emission of MAMMOTH-1 arises mainly from the diffuse nebula rather than from the point-spread function (PSF). In Figure~4, we present the radial profile of MAMMOTH-1's surface brightness. This ELAN has an extremely high extended nebular luminosity, and the central PSF contributes only $4\%$ of the total Ly$\alpha$ luminosity. 
If we subtract the Ly$\alpha$ PSF (source B in Figure~1), 
MAMMOTH-1 has an extended nebular Ly$\alpha$ luminosity of $5.07\pm0.07\times10^{44}$ erg s$^{-1}$,  the highest discovered to date. We summarize the size and luminosity of MAMMOTH-1 in  Table~1. 

In Figure~2,  the northern/eastern part of MAMMOTH-1 seems to have a filamentary structure. If this filamentary structure is real, it aligns the same direction with the morphology of the large-scale structures (see Cai et al. 2016a). In the cosmic hierarchical nature of structure formation, large-scale filaments are formed out of the merging of small-scale pieces. Simulations suggest that 
 cosmic webs containing baryonic matter tend to align with underlying large-scale structures of dark matter (e.g. Cen et al. (1994); Cen \& Ostriker (2006); Fukugita \& Peebles (2004), Colberg et al. 2005, Hellwing 2014). Our observations tentatively support these simulations.

\figurenum{1}
\begin{figure*}[tbp]
\epsscale{1.0}
\label{fig:02+04}
\plotone{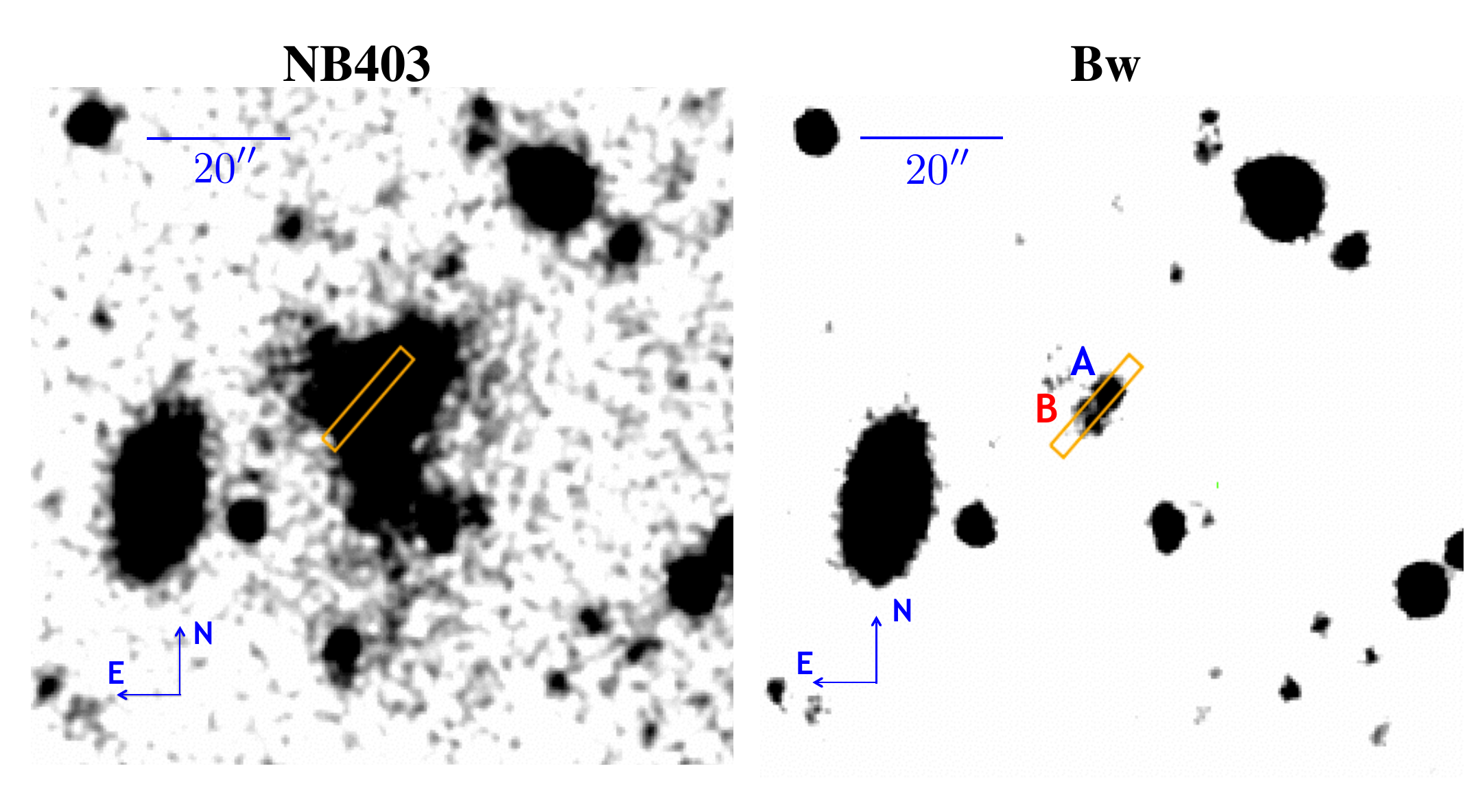}
\caption{Final stacked images of the field surrounding MAMMOTH-1 overplotted with our LBT/MODS slit (orange rectangle). Each image has an area of $80''$ centered on the Ly$\alpha$ nebula. The narrow-band (NB403) image (left panel) at the Ly$\alpha$ line of the redshift of MAMMOTH-1 ($z=2.32$). The deep Bw-band image (middle panel) does not show any extended emission associated with MAMMOTH-1. We put the spectral slit (orange slit) containing the brightest two sources (A, B) in the center of MAMMOTH-1. }
\end{figure*}

\figurenum{2}
\begin{figure*}[tbp]
\epsscale{1.2}
\label{fig:02+04}
\plotone{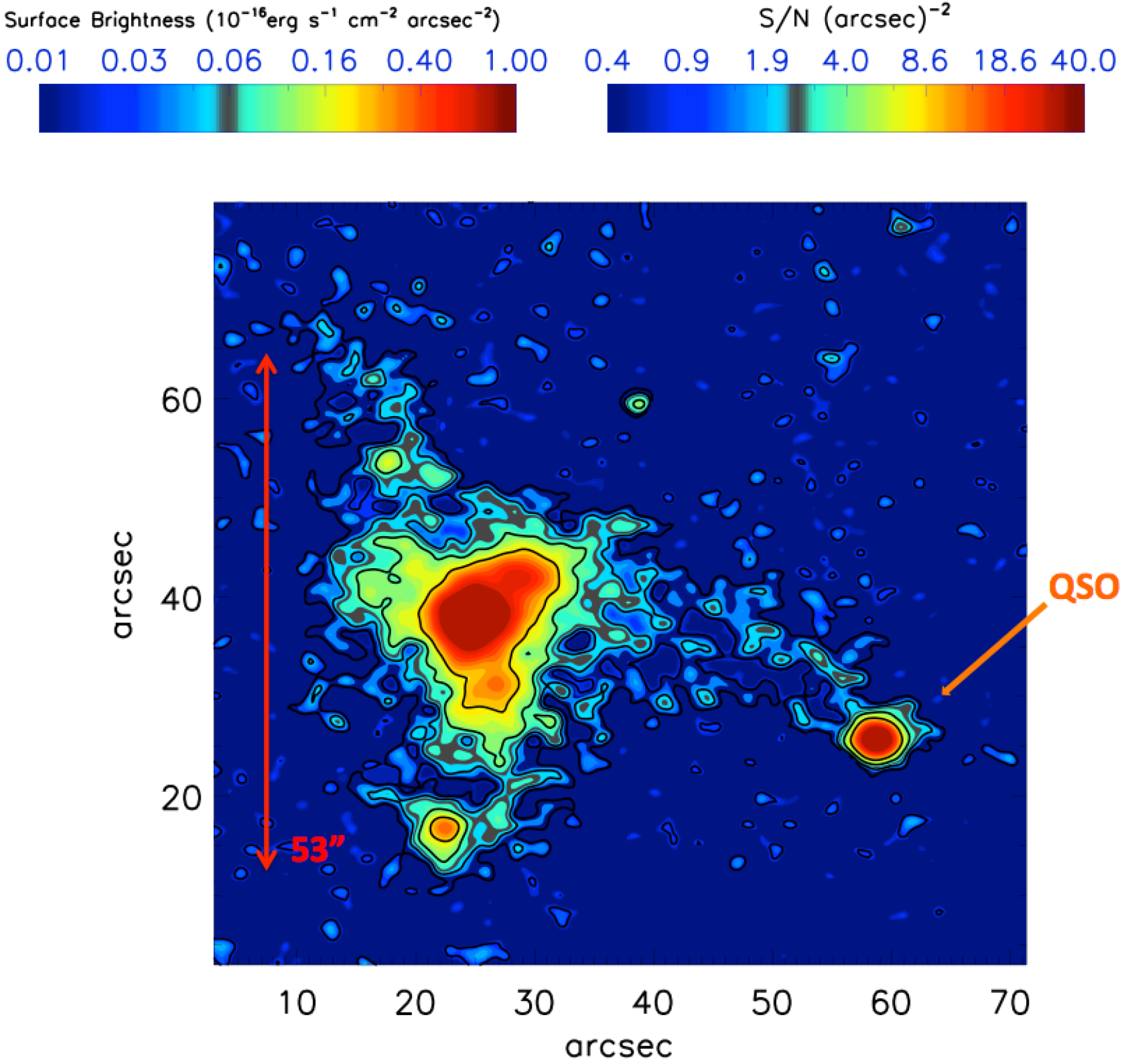}
\caption{Continuum-subtracted, smoothed narrow-band image of the field around the enormous Ly$\alpha$ nebula  (ELAN) MAMMOTH-1. The color map and contours indicate the Ly$\alpha$ surface brightness (left color bar) and the signal-to-noise ratio per arcsec$^2$ 
aperture (right color bar), respectively. This image reveals the Ly$\alpha$ emission of the enormous Ly$\alpha$ nebula (ELAN). The current 1$\sigma$ surface brightness limit is $2.4\times10^{-18}$ erg s$^{-1}$ cm $^{-2}$ arcsec$^{-2}$. Above the flux contour of   $SB>4.8\times10^{-18}$ erg s$^{-1}$ cm$^{-2}$ arcsec$^{-2}$, MAMMOTH-1 has a total luminosity of $L= 5.28\pm0.07\times10^{44}$ erg s$^{-1}$. Further, we tentatively detect filamentary structures around MAMMOTH-1. }
\end{figure*}

\figurenum{3}
\begin{figure}[tbp]
\includegraphics[width=0.52\textwidth,height=0.28\textheight]{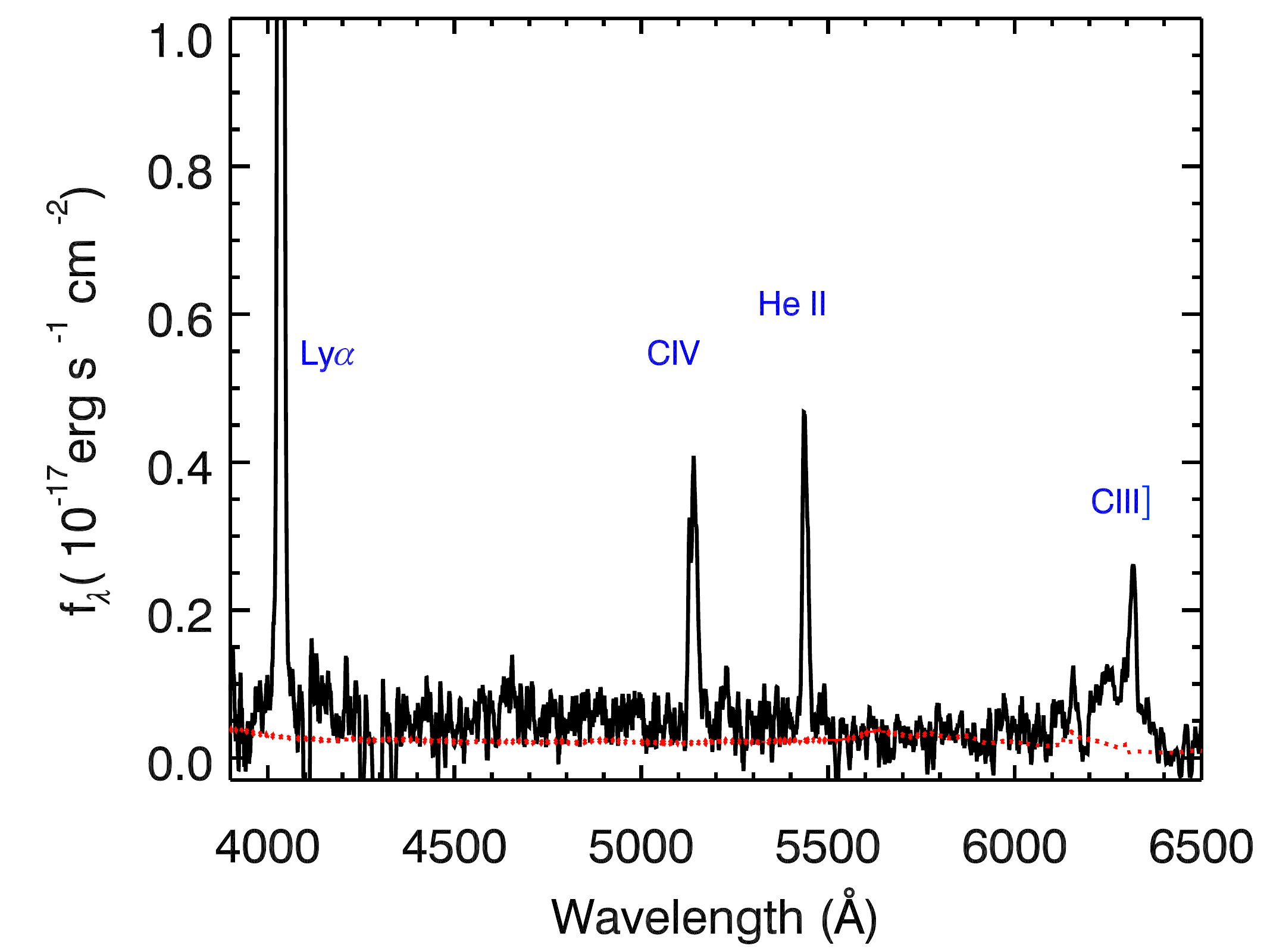}
\caption{ The LBT/MODS spectrum of the enormous Ly$\alpha$ nebula (ELAN) MAMMOTH-1 at $z=2.32$ centered on source B in Figure 1. This spectrum is taken using a long-slit with a 2$''$ slit width. The red dotted line represents the error of the spectrum.} 
\end{figure}

\figurenum{4}
\begin{figure}[tbp]
\epsscale{1}
\label{fig:02+04}
\plotone{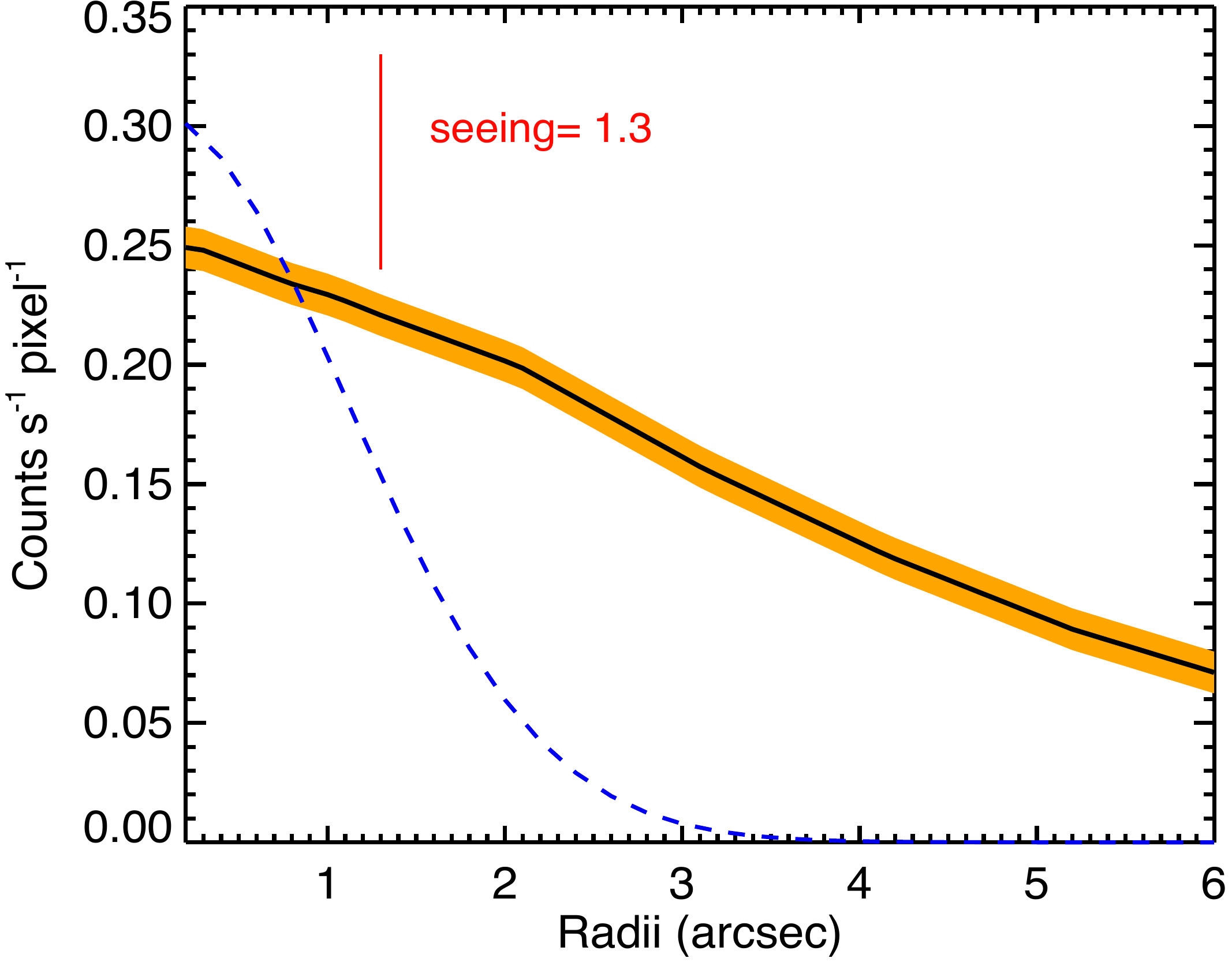}
\caption{Surface brightness as a function of radii. MAMMOTH-1 has a very extended profile (black line). Orange present the 1-$\sigma$ error of the radial profile. We use source B's position (Figure 1) as the center of MAMMOTH-1. Blue dashed line represents central point-spread-function (PSF). The PSF is constructed using a full-width-half-maximum (FWHM) that is equal to the seeing of 1.3$"$. The amplitutde is determined assuming all of the ELAN's luminosity within the Moffat FWHM is contributed by the PSF.  The central PSF constitutes only 4\% of entire flux of MAMMOTH-1.  }
\end{figure}

\vspace{0.05in}

{\subsection{Emission Line Profiles}}

The deep LBT/MODS spectra  reveal Ly$\alpha$, \heii, \civ, and \ciii\ emission (Figure 4). Both \civ\  
and \heii\ extend over $\gtrsim 30$ kpc scales (Figure 5). Extended \heii\ and \civ\ have been  observed previously in radio galaxies, but MAMMOTH-1 is unlikely to be powered by radio jet. From the FIRST radio catalog (Becker et al. 1995), we do not find any radio-loud sources with a radio flux at 1.4 GHz  $F({\rm{1.4GHz}})>0.9$ mJy in the area within 30 arcsec of the MAMMOTH-1 nebula. 
We use the redshift of the non-resonant \heii\  $\lambda1640$ line as the redshift of MAMMOTH-1, yielding $z=2.319\pm0.004$.


The Ly$\alpha$, \civ, and \heii\ line profiles reveal two main components. In Figure 5, we fit these lines with two Gaussians. For the Ly$\alpha$ line,  the  blue component has a best-fit FWHM of $876\pm120$ km s$^{-1}$ and the red  component has a best-fit FWHM of $1140\pm160$ km s$^{-1}$. The redshift offset between the two components is $\approx 700$ km s$^{-1}$. For the \heii\ line, the blue component has a FWHM of $714\pm100$ km s$^{-1}$, and the red component has a best-fit FWHM of $909\pm130$ km s$^{-1}$. The offset between the two components is the same as that of Ly$\alpha$.

\figurenum{5}
\begin{figure*}[tbp]
\epsscale{1.0}
\plotone{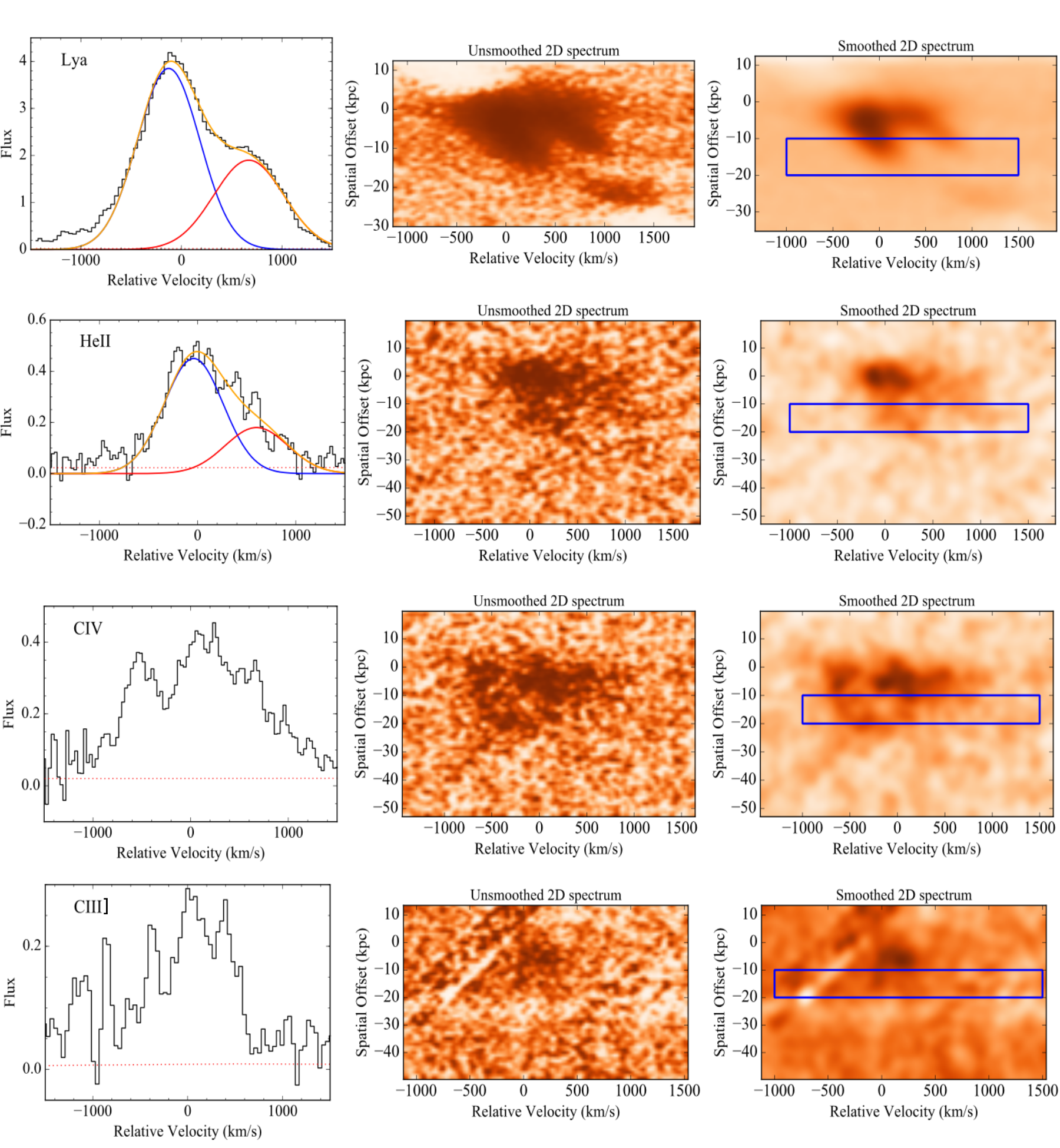}
\caption{The spectrum of emission lines of MAMMOTH-1 observed in LBT/MODS. From top to bottm: Ly$\alpha$, \heii, \civ, and \ciii. {\it Left column:} 1-D zoom of line profiles of Ly$\alpha$,  \heii,  \civ, and \ciii\ (from top to bottom) as a function of the rest-frame velocity centered on the redshift of the nebula, measured from the \heii\ line. We use two Gaussian functions to fit the two major components of Ly$\alpha$ and \heii.  Our spectral resolution is $R\approx1000$ for this slit (2$''$ wide), corresponding to $91$ km s$^{-1}$ in the rest-frame of MAMMOTH-1. The measured FWHM for each component is much wider than the spectral resolution.  The deviation of \ciii\ emission in the left wing is due to imperfect sky subtraction. {\it Middle column: } The unsmoothed 2D spectra of Ly$\alpha$, \heii, \civ, and \ciii. {\it Right column:} The $2\times2$-pixel smoothed spectra of Ly$\alpha$, \heii, \civ, and \ciii. 
From the spectrum, the extended Ly$\alpha$ emission lies the spatial direction along the slit ($10"$). The \civ\ extends 33 physical kpc,  and \heii\ extends 31 physical kpc. The \ciii\ is on a much smaller scale than \civ\ and \heii. We show the aperture for measuring the surface brightness (blue rectangles). }
\end{figure*}

\vspace{0.05in}

\subsection{Flux and Surface Brightness}

In our LBT/MODS spectra, the slit is 10$''$ long and 2$''$ wide. We measure the flux of Ly$\alpha$, \heii, \civ, and \ciii. The aperture we applied is 15 $\pm5$ kpc (1.8$\pm0.6''$) away from source B along the slit direction, and within 3000 km s$^{-1}$ in the wavelength direction (see blue rectangle in Figure~5), sufficiently large to include all the diffuse emission in the wavelength direction. This gives a size of 2$''$ (slit width) $\times1.2''$ (along the slit direction). We apply this aperture to measure the surface brightness of the emission lines. We regard the flux as the CGM emission at $R=$ 1.8$''$ (15 kpc) away from the central  source.  

Applying this aperture to the LBT/MODS 2D spectrum (Figure~5), we determine that the Ly$\alpha$ emission  (first panel) has a flux of $f_{\rm{Ly\alpha},15kpc}=6.6\pm0.2 \times10^{-16}$ erg s$^{-1}$ cm$^{-2}$, corresponding to $L_{\rm{Ly\alpha},15kpc}=1.7\times10^{43}$ erg s$^{-1}$. The surface brightness of Ly$\alpha$ is SB$_{\rm{Ly\alpha,15kpc}}= 2.99\pm0.01 \times10^{-16}$ erg s$^{-1}$ cm$^{-2}$ arcsec$^{-2}$.

The extended \heii\ emission (second panel of Figure~5) has a line flux of $f_{\rm{HeII, 15kpc}}= 7.8\pm0.2\times10^{-17}$ erg s$^{-1}$ cm$^{-2}$, corresponding to a luminosity of $L_{\rm{HeII, 15kpc}}= 3.2\pm0.1 \times10^{42}$ erg s$^{-1}$, and a surface brightness $SB_{\rm{HeII, 15kpc}}= 3.3\pm0.1\times10^{-17}$ erg s$^{-1}$ cm$^{-2}$ arcsec$^{-2}$.

 The extended \civ\ emission (third panel of Figure~5) has a line flux of $f_{\rm{CIV, 15kpc}}= 8.8\pm0.2\times10^{-17}$ erg s$^{-1}$ cm$^{-2}$, corresponding to $L_{\rm{CIV, 15kpc}}= 3.6\pm0.1 \times10^{42}$ erg s$^{-1}$ and $SB_{\rm{CIV, 15kpc}}= 3.7\pm0.1\times10^{-17}$ erg s$^{-1}$ cm$^{-2}$ arcsec$^{-2}$. 

The  \ciii\ emission (fourth panel of Figure~5) has a line flux of $f_{\rm{HeII, 15kpc}}= 0.9\pm0.2\times10^{-17}$ erg s$^{-1}$ cm$^{-2}$, corresponding to  $L_{\rm{HeII, 15kpc}}= 0.4\pm0.1 \times10^{42}$ erg s$^{-1}$ and  $SB_{\rm{HeII, 15kpc}}= 0.4\pm0.1\times10^{-17}$ erg s$^{-1}$ cm$^{-2}$ arcsec$^{-2}$. 
{ We summarize our surface brightness measurements in Table~2. }

{\subsection{Comparison between MAMMOTH-1 and Other Ly$\alpha$ Nebulae}}

In Figure~6, we present the sizes and Ly$\alpha$ luminosities for different Ly$\alpha$ nebulae from the literature in comparison with MAMMOTH-1. 
The Ly$\alpha$ emission from the central source has also been included. The  typical size measurements for these objects are above surface brightness contours of $\sim 5\times 10^{-18}$ erg s$^{-1}$ cm$^{-2}$ arcsec$^{-2}$, comparable to our measurements for MAMMOTH-1, which are above $4.8\times10^{-18}$ erg s$^{-1}$ cm$^{-2}$ arcsec$^{-2}$. If we restrict the size measurements to the surface brightness coutour of $4.8\times10^{-18}$ erg s$^{-1}$ cm$^{-2}$ arcsec$^{-2}$, MAMMOTH-1 has a similar size to the Slug nebula (Cantalupo et al. 2014). But note that \citet{cantalupo14} reached a factor of $3 \times$ deeper than our current narrowband imaging, so MAMMOTH-1 nebula may be extended on an even larger scale in deeper data. 
In Table~3, we further compare the diffuse nebular luminosities (with excluding the PSF contribution). MAMMOTH-1 has the highest diffuse nebular luminosity among all the confirmed Ly$\alpha$ nebulae and ELANe.  


\figurenum{6}
\begin{figure}[tbp]
\epsscale{1.2}
\label{fig:02+04}
\plotone{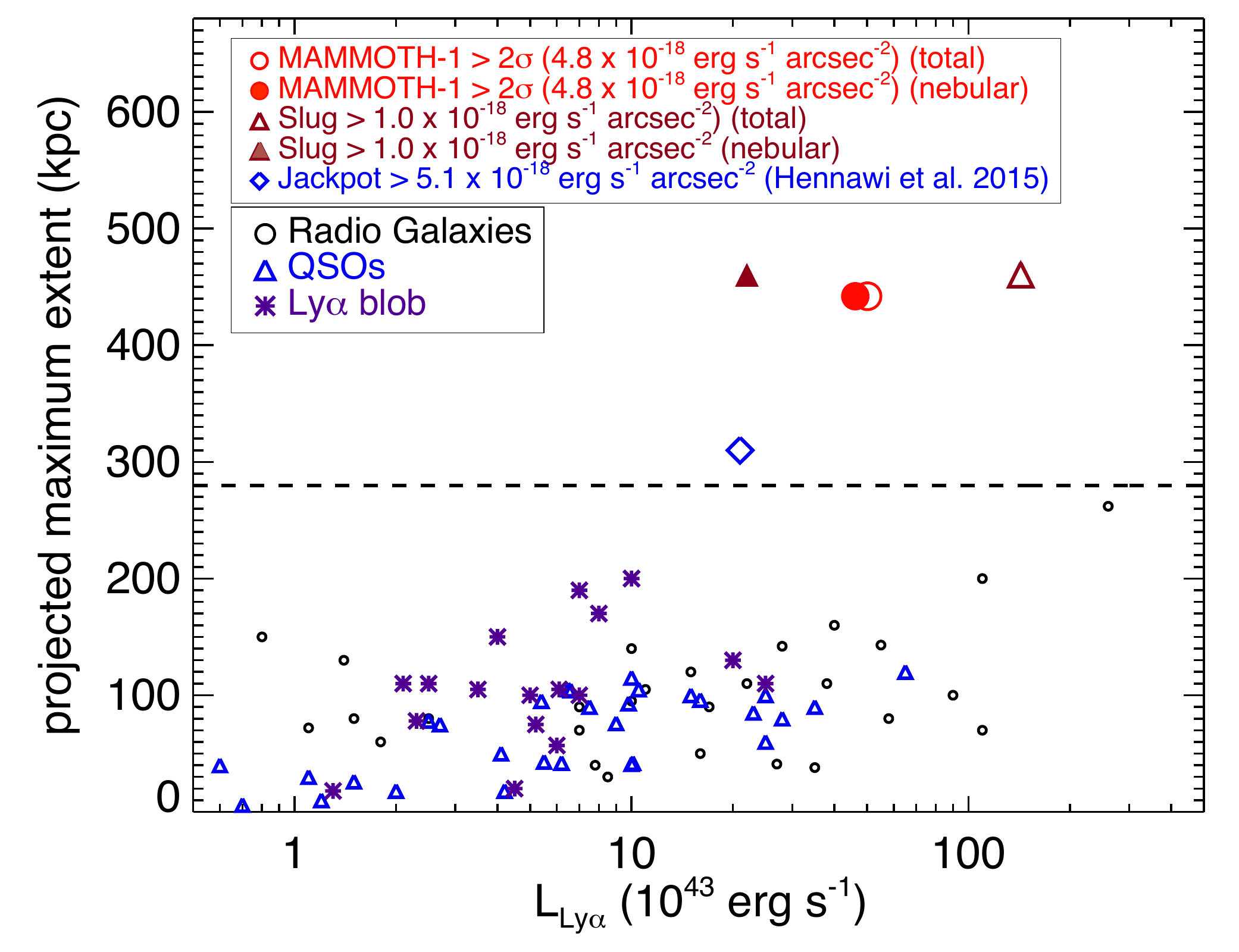}
\caption{Projected maximum extent versus total Ly$\alpha$ luminosity for different objects from the literature. The typical size measurement is above a surface brightness contour of $5\times 10^{-18}$ erg s$^{-1}$ cm$^{-2}$ arcsec$^{-2}$. If we restrict the size measurement to contours above $4.8\times10^{-18}$ erg s$^{-1}$ cm$^{-2}$ arcsec$^{-2}$, our target MAMMOTH-1 is one of the  most extended sources, with a comparable spatial extent to the Slug nebula. The open circle represents the total Ly$\alpha$ luminosity of MAMMOTH-1 and the filled circle represents the nebular luminosity, with excluding the contribution from the central point spread function (PSF). The luminosities of Slug nebula are cited from Cantalupo et al. (2014). The black dashed line shows the virial diameter of a dark matter halo of mass $M \sim 10^{12.5}\ M_\odot$, the typical host of radio-quiet QSOs (Martin et al. 2012, Cantalupo et al. 2014). }
\end{figure}

\section{Discussion}

In the previous section, we showed that the ELAN MAMMOTH-1 has a Ly$\alpha$ spatial extent of  $\approx440$ kpc and a total luminosity of $5.28\pm0.07\times10^{44}$ erg s$^{-1}$. This nebula resides in an extremely overdense galaxy environment previously discovered at $z=2.3$. Moreover, this radio-quiet 
nebula has the strongly extended \heii, \civ, and \ciii emission (Figure~4). The Ly$\alpha$, \heii, \civ\ line profiles are all double-peaked. In Table~3, we compare the properties of MAMMOTH-1 to other ELANe recently discovered. 
The Ly$\alpha$ spatial extent and the strong emission of \civ\ and \heii\ make MAMMOTH-1 unique. 
In this section, we derive the physical properties of  MAMMOTH-1, and 
we discuss several possible physical explanations for powering this ELAN.

\subsection{Ionizing Radiation}

 A comparison between hydrogen  ionizing  photons and helium ionizing  photons constrains the hardness of the ionizing radiation. The number of H$^+$ ionizing photons can be expressed as: 

\begin{equation}
Q(\rm{H})= \frac{L_{\rm{Ly\alpha, 15kpc}}}{h\nu_{\rm{Ly\alpha}}}\frac{1}{0.68}\approx 1.5\times10^{54}\ \rm{s}^{-1}
\end{equation}
where $f_{\rm{Ly\alpha},15kpc}=2.99\pm0.01\times10^{-16}$ erg s$^{-1}$ cm$^{-2}$, corresponding to $L_{\rm{Ly\alpha},15kpc}=1.7\times10^{43}$ erg s$^{-1}$. We have assumed that  $\approx 68$\% of the ionizing photons are converted into Ly$\alpha$ emission \citep{spitzer78}. This is a lower limit of Ly$\alpha$, because it may be destroyed by dust. 

Using the same spatial region, we measured that the \heii\ emission has a flux of $f_{\rm{HeII}}=7.8\pm0.2\times10^{-17} $ erg s$^{-1}$ ($L_{\rm{HeII}} = 3.2\pm0.1\times10^{42}$
erg s$^{-1}$). We calculated the He$^+$-ionizing photon number ($E_{\nu} \ge54.4$ eV) using the equation of:
\begin{equation}
Q(\rm{He^+})= \frac{L_{\rm{\lambda1640}}}{h\nu_{\rm{\lambda1640}}}\frac{\alpha^{\rm{eff}}_{HeII}}{\alpha^{1640}_{\rm{HeII}}}\approx 2.8\times10^{53}\ \rm{s}^{-1}
\end{equation}
where we assumed the case B recombination model, with a temperature of $T=10^4$ K. 
Under this assumption, $\alpha_{\rm{HeII}}^{\rm{eff}}(T)= 1.53\times10^{-12}$ cm$^3$ s$^{-1}$ (Prescott et al. 2009). Therefore  Q(He$^+$)/Q(H) is equal to 0.19. Note this is an upper limit for the Q(He$^+$)/Q(H) ratio because Ly$\alpha$ could be destroyed by dust. The Q(He$^+$)/Q(H) ratio suggests that the powering source of  MAMMOTH-1 produces a hard ionizing radiation spectrum.  
In comparison, if we assume a typical Pop II stellar population with a Salpeter IMF, and a low metallicity of $Z=10^{-4}\ Z_\odot$, 
then Q(He$^+$)/Q(H) equals 0.005 \citep{schaerer03, prescott09}, two orders of magnitude smaller than the value we estimated 
from MAMMOTH-1. This hard ionization ratio could arise because of significantly lower metallicity (e.g., Population III), a stellar population with a top-heavy IMF \citep{tumlinson03, schaerer03, cai11}, or an AGN. 
The detection of strong \civ\ and \ciii\ emission make this nebula unlikely to be powered by a low metallicity (e.g., Pop III) stars. Our current data support the conclusion that this ELAN is powered by one or more hard ionizing sources (e.g., AGN).

\subsection{Sources of the enormous Ly$\alpha$, strong extended \civ, and \heii\ emission in a radio-quiet system}

At least 15 Ly$\alpha$ nebulae with Ly$\alpha$ spatial extents larger than $150$ kpc have recently been discovered \cite[e.g.,][]{cantalupo14, hennawi15, borisova16}. But in none of these nebulae have strongly extended \heii\ and \civ\ been reported. We will discuss below several mechanisms that may power MAMMOTH-1.

\subsubsection{Photoionization Model}

In photoionization models, \civ\ emission is mainly powered by collisional excitation \cite[e.g.,][]{arrigoni15a}. The intensity of  
collisional excitation has a strong dependence on the temperature \cite[e.g.,][]{gurzadyan97}. A higher ionization parameter ($U$) yields a higher gas temperature, and thus the \civ\ intensity strongly depends on the ionization parameter. Collisional excitation also depends on the gas density and column density of \civ. The \heii\ emission is mainly due to  recombination. The fraction of \heii\ emission reaches a peak at $U\sim-2.0$, where a larger fraction of the helium  has been doubly ionized (e.g., Arrigoni-Batta et al. 2015b).  Higher ionization parameters only modestly change the \heii\ intensity.   The \ciii\ emission increases with the ionization parameter, and it is also highly sensitive to the metallicity. The \ciii\ emission peaks at a gas metallicity of $Z \sim 0.2\times Z_\odot$, and it decreases at both higher and lower metallicities (Erb et al. 2009). Therefore, the combination of \heii, \civ, and \ciii\ strongly constrains the physical properties of the CGM. 

Using CLOUDY ionization modeling \citep{ferland96}, \citet{arrigoni15a} have thoroughly investigated the \heii/ Ly$\alpha$ and \civ/Ly$\alpha$ ratios under different ionization parameters, gas densities ($n_{\rm{H}}$), and QSO ionizing luminosities ($L_{\nu_{{\rm{LL}}}}$) for the Slug nebula and nebulae in SSA22 protocluster.
In \S3, we suggest that MAMMOTH-1 could be powered mainly by source B. Source B may be a strongly obscured source, e.g., a type-II AGN. The Ly$\alpha$ emission from a strongly obscured source may be complicated to interpret. In this section, we conduct a similar studies as \citet{arrigoni15a}, but focus on reproducing the \heii\ surface brightness and the \civ/\heii\ and \ciii/\heii\ line ratios. In our CLOUDY modeling, the AGN continuum follows the recipe in Mathews \& Ferland (1987). We assume that the CGM clouds have a constant hydrogen density ($n_{\rm{H}}$). We assume that the emitting gaseous clouds are uniformly distributed throughout the halo, and we further assume a standard plane-parallel geometry for these clouds. To match our measurements in \S3.3, we assume that the distance between the CGM cloud and the central QSO is $R\approx15$ kpc.  In our CLOUDY models, we try combinations of different $n_{\rm{H}}$ values, with $n_{\rm{H}}= 0.01 - 10.0$ cm$^{-3}$ (steps of 0.5 dex);  different ionization parameters, with Log $U= -3  -1$ (steps of 0.5 dex); different column densities of $N_{\rm{H}}=10^{19} - 10^{22}$ cm$^{-2}$ (steps of 0.5 dex), and metallicities with $Z=0.1 - 1.0\times Z_\odot$ (steps of $0.5\times Z_\odot$). We assume a gas covering fraction of $f_C=0.3$ \cite[e.g.,][]{cantalupo14}. 

Our observed \heii\ surface brightness is $SB_{\rm{HeII}}\approx3.3\times10^{-17}$ erg s$^{-1}$ cm$^{-2}$ arcsec$^{-2}$. We require that the parameter combinations yield a \heii\ surface brightness of $\approx 3.0-3.5\times10^{-17}$ erg s$^{-1}$ cm$^{-2}$ arcsec$^{-2}$ to roughly match the observed \heii\ emission. In Figure~7,  we present models that yield the observed \heii\ surface brightness, and present our observed value using a red dot with an error bar. Using the parameter combinations with ($N_{\rm{H}}$, $Z$, Log(U), $n_{\rm{H}}$)= (10$^{20}$ cm$^{-2}$, 0.5 $Z_\odot$, $-2$, 0.1 cm$^{-3}$) and ($10^{18}$ cm$^{-2}$, 1.0 $Z_\odot$, $-2$, 2.0 cm$^{-3}$) reproduce the observed intensities of \heii, \civ\ and \ciii, and the line ratios of \civ/\heii\ and \ciii/\heii\ within 1$\sigma$ errors (red error bars in Figure~7). Therefore, the \civ/\heii\ and \ciii/\heii\ line ratios are consistent with AGN photoionization. 


\figurenum{7}
\begin{figure}[tbp]
\epsscale{1.2}
\plotone{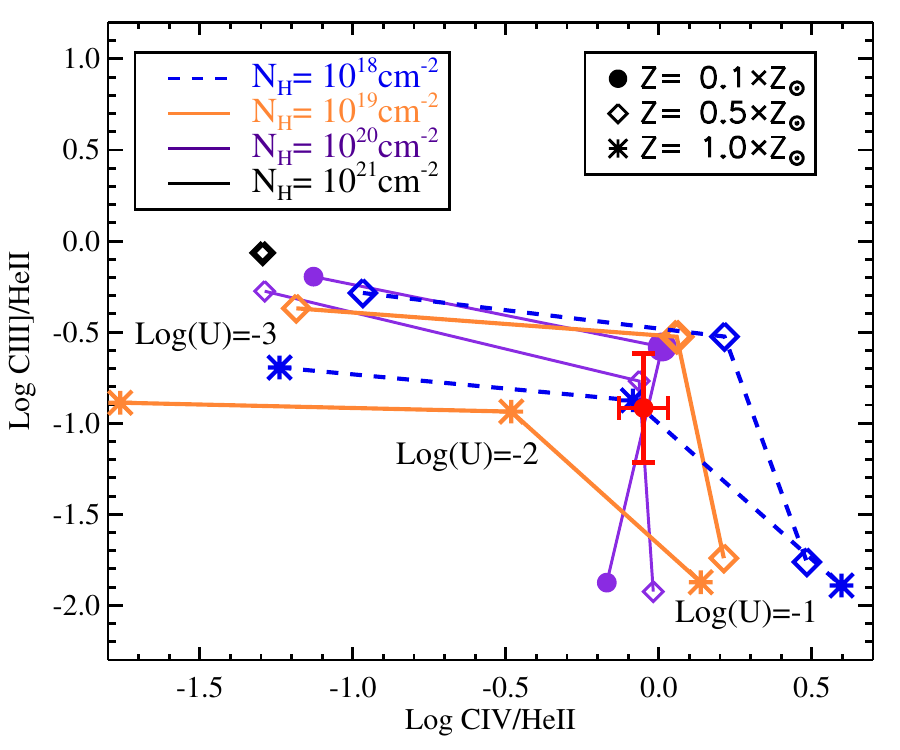}
\caption{Simulation of the intensity ratios of \civ/\heii\ and \ciii/\heii\ powered by AGN. Colors represent different ionization parameters ($LogU$) and symbols represent different metallicities of the gaseous clouds. The observed value is marked with red points with an error bar. MAMMOTH-1 is consistent with the photoionization scenario, with an ionization parameter of Log U $\approx 2$ and a gas metallicity of 0.1 $Z_\odot$.}
\end{figure}

\subsubsection{\it Resonant Scatter}

In \S3, we have shown that the Ly$\alpha$, \civ\ and \heii\ emission line profiles contain at least two major components. 
Double-peaked Ly$\alpha$ emission is predicted by the resonant scattering of Ly$\alpha$ photons \cite[e.g.][]{dijkstra06, yang14}. The key prediction of these radiative transfer (RT) calculations is that the Ly$\alpha$ spectrum is double peaked with an enhanced blue peak, producing a blueshift of the Ly$\alpha$ profile. Although it is true that this prediction matches the Ly$\alpha$ profile of MAMMOTH-1, the \heii\ emission has the same double-peaked structure as Ly$\alpha$. \heii\ is an optically-thin, non-resonant emission line whose photons escape the halo without radiative transfer effects \cite[e.g.,][]{yang14}. 
The non-resonant, optically thin emission lines should directly reflect the spatial distribution and kinematics of the gas. Thus, the emission-line structure of \heii\ strongly suggests that the double peaks are due to the actual kinematics (e.g., gas flows) rather than the radiative transfer effects.

\subsubsection{\it Shocks due to a Gas Flow }

The shocks due to flowing gas can also explain the double peaks of the emission lines.  If the fast wind of an outflow is launched, then the shock could heat the surrounding interstellar gas over scales of $\gtrsim 50$ kpc (\citealt{debuhr12, harrison14}).  Current galaxy formation simulations and observations suggest that high-velocity ($v_{\rm{max}}\sim1000$ km s$^{-1}$) galactic outflow can quench star formation in the most massive
galaxies and eject heavy elements into the IGM \cite[e.g.,][]{taniguchi00, martin05, ho14}.  
 Such galactic winds can be driven by (1) intense star formation or (2) relativistic winds or jets resulting from the gas accretion onto the supermassive black holes \cite[e.g.,][]{leitherer99, tombesi15}. 

 \citet{wilman05} find  a Ly$\alpha$ blob at $z=3.09$ in the SSA22 overdensity \citep{steidel00, matsuda05} whose double-peaked line profile is consistent with a simple outflow model. This model suggests that the Ly$\alpha$ emission is absorbed by a foreground shell of neutral gas that is pushed out up to a $\approx70$ kpc by an AGN-driven outflow.  
Using MAPPINGS \citep{dopita96} and CLOUDY \citep{ferland96} modeling, \citet{villar99, villar07} and \citet{moy02}  suggest both shocks and AGN photoionization could power the extended Ly$\alpha$, \heii, and \civ\ observed in radio galaxies.  
 Using hydrodynamical simulations, \citet{cabot16} further argue  that the Ly$\alpha$, \heii\ and \civ\ emission in $z\approx3$ Ly$\alpha$ blobs could be primarily due to the shocks. Integral Field Spectrometer (IFS) observations suggest that the high-velocity ($v_{\rm{max}}\approx1000$ km s$^{-1}$) [OIII] outflows exist in a sample of 5 radio-quiet ULIRGs at $z\gtrsim2$. Such [OIII] outflows are consistent with the AGN-driven wind scenario \cite[e.g.,][]{alexander10, harrison12}. 

MAMMOTH-1 has \ciii/\heii\ and \civ/\heii\ line ratios consistent with both photionization and shock models (see Figure~2 and Figure~3 of \citet{villar99}. Further, the \civ/Ly$\alpha$ and \heii/Ly$\alpha$ ratios of MAMMOTH-1 are consistent with the predictions using shock models \citep{arrigoni15a}, with a gas denisty $n_H\sim 0.1-$ 1 cm$^{-3}$ and a shock velocity of $500$ -- 600 km s$^{-1}$. If the extended \heii\ and \civ\ are powered by shocks due to an AGN-driven outflow, then the  double velocity peaks of emission lines can be naturally interpreted. Like \citet{harrison14}, we draw a schematic diagram to illustrate the outflow interpretation of the extended \civ\ and \heii\ (Figure~8). The velocity offsets between the two components and the spatial extent of emission lines strongly depend on the orientation of the outflow with respect to the line of sight:  if the axis of the outflow is oriented along the line of sight, then a high-velocity offset and a small spatial extent should be observed; and conversely, if the axis of the outflow is in the plane of the sky, then a small velocity offset and a large spatial extent should be observed.


 From \S3.2,  the offset between two velocity components is $\approx700$ km s$^{-1}$. These line structures are similar to ULIRG sample in \citet{harrison12}. The AGN outflow is a natural explanation of the extended metal line emission. If we assume the extended \civ\ and \heii\ are due to the AGN outflow, { then we can estimate the energy of the outflow.  
 
 To make an order of magnitude estimation of the outflow energy, we follow the model proposed by Heckman et al. (1990) (also see Nesvadba et al. 2008; Harrison et al. 2012). This model assumes a energy-conserving bubble  inflated by AGN outflow  expanding into a uniform circumgalacitc medium (CGM) with a constant density. Further, this model assumes that the outflow injects energy to CGM at a constant rate. Then, Heckman et al. (1990) derived the following equation to calculate the injected energy coupled to the CGM (E$_{\rm{kin}}$): }
\begin{equation}
\dot{E_{\rm{kin}}}\approx 1.5\times10^{46}\ r_{10}^2\ v^3_{1000}\ n_{0.5} \ \rm{erg}\ \rm{s}^{-1}
\end{equation}
where $v_{1000}$ is the velocity offset between two components in units of 1000 km s$^{-1}$. $r_{10}$ is the radius of the observed of \civ\ emission in units of 10 kpc. 
The ambient density is the gas density ahead of the expanding bubble, in the units of 0.5 cm$^{-3}$. { The total energy injected is $E_{\rm{in}}= E_{\rm{kin}}/\eta$, where $\eta$ is the coupling efficiency. The coupling efficiency represents the fraction of the outflow energy that is coupled to circumgalactic medium (CGM) gas. The value of $\eta$ could range from 0.05 -- 0.8 (Nesvadba et al. 2008). For MAMMOTH-1, if we assume that the extended \civ\ and \heii\ are completely powered by an AGN outflow, and the axis of the outflow is oriented 45 degrees with respect to the sight line, then $r_{10}=2$, $v_{1000}=0.7$.  It is actually hard to find a direct way of estimating the electron density $n_{0.5}$.  We assume the electron density $n_{0.5}=$ 1 -- 4 which values were indirectly measured from Heckman et al. (1990) and was dopted by Heckman et al. (1990), Nesvadba et al. (2008) and Harrison et al. (2012) when they calculating the AGN feedback energy at $z=2-3$. Taking these numbers into Equation (3), then we make an order of magnitude estimation that the outflow in MAMMOTH-1 injects energy into the CGM at a considerable rate of $\sim 10^{45-47}$ erg s$^{-1}$. } 
Over a typical AGN duty cycle of 30 Myr \cite[e.g.,][]{hopkins05, harrison12}, the total energy injected reaches the order of $10^{60- 62}$ erg. According to \citet{nesvadba06}, the typical binding energy of a massive elliptical galaxy with a halo mass of  $M_{\rm{halo}}\approx10^{12}$ M$_\odot$ is about $10^{60}$ erg.  Thus, if MAMMOTH-1 is powered by an AGN outflow, then the outflow energy could be comparable or two orders of magnitude higher than this binding energy, making a vast AGN outflow possibly plays a major role in heating the ISM.



It has also long been suggested that jet-induced shocks can power extended metal-line emission, and extended \civ\ emission has been reported in a few radio-galaxies with strong radio continua \cite[e.g.,][]{mccarthy93, villar07}. We argue that our current data disfavour the model of  jet-ISM interaction. From the FIRST radio catalog (Becker et al. 1995), we do not find any source with a radio flux at 1.4 GHz $>0.9\mu$Jy within a radius of 30 arcsec from MAMMOTH-1. Assuming a radio spectrum $S(\nu)\propto \nu^{-0.8}$, this 3-$\sigma$ upper limit corresponds to a luminosity density of $< 3.2 \times 10^{32}$ erg s$^{-1}$ Hz$^{-1}$ at rest-frame 1.4 GHz \citep{yang09}. This limit is two orders of magnitude lower than the radio continua of other Ly$\alpha$ nebulae powered by radio galaxies \cite[e.g.,][]{carilli97,reuland03}.

\figurenum{8}
\begin{figure}[tbp]
\epsscale{1}
\plotone{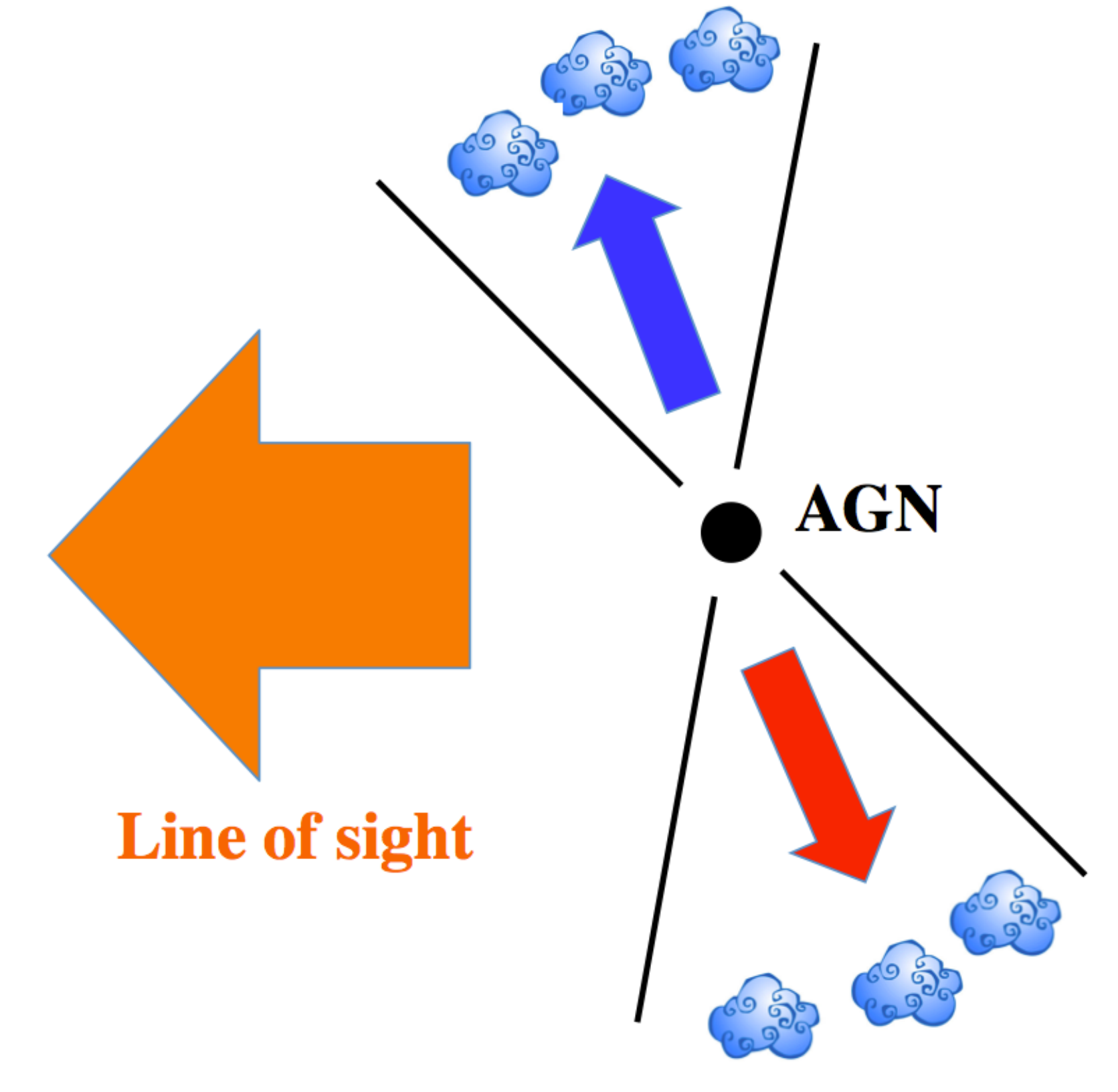}
\caption{A schematic diagram to demonstrate the outflow interpretation of the data. The \civ\ and \heii\ velocity offsets between the two components and the spatial extent of emission lines shown in Figure~5 strongly depend on the orientation of the outflow with respect to the line of sight. For a given AGN outflow, if the axis of the outflow is oriented along the line of sight, high-velocity offsets and a small spatial extent would be observed. Conversely, if the axis of the outflow is in the plane of the sky, a small velocity offset and a large spatial extent would be seen. }
\end{figure}

\subsubsection{\it Gravitational Cooling Radiation}

 Theoretical studies have suggested that Ly$\alpha$ nebula could result from the gravitational cooling radiation \cite[e.g.,][]{haiman01, dijkstra06, yang06, faucher10, rosdahl12}.  
Several studies have predicted the \heii\ cooling radiation using hydrodynamical simulations. 
 \citet{yang06} predict that the \heii\ line has the FWHM $\le 400$ km s$^{-1}$ even for the most massive halo at $z\approx2$ ($M\sim10^{14}\ M_\odot$).  If our observed \heii\ line profile has two major velocity components as shown in Figure~5, then the \heii\ has a large FWHM of $714\pm100$ km s$^{-1}$ for the blue component and $909\pm130$ km s$^{-1}$ for the red component. The observed FWHMs are much wider than the predicted line width for cooling radiation. 
Also, using hydrodynamical simulations, \citet{fardal01} and \citet{yang06} point out that the \heii\ regions should be centrally-concentrated and the \heii\ cooling radiation  may be too small to resolve using current ground-based telescopes. This size prediction of the \heii\ cooling radiation does not fit our observations. We have detected  extended \heii\ emission over $\gtrsim30$ kpc scale. 
Further, if the Ly$\alpha$ emission results from the cooling inflow of the pristine gas in the intergalactic filaments, then we should expect no extended \civ\ being detected \cite[e.g.,][]{yang06, arrigoni15a}. 
Therefore, we conclude that our current observations do not fit with the cooling radiation picture.

\section{Summary}

In this paper, we present our discovery of an enormous Ly$\alpha$ nebula (ELAN) MAMMOTH-1 at $z=2.319$ in the density peak of the large-scale structure BOSS1441 (Cai et al. 2016a). 
Above the 2$\sigma$ surface brightness contour, this object has the highest nebular luminosity discovered to date:  $L_{\rm{Ly\alpha}}=5.1\pm0.1\times10^{44}$ erg s$^{-1}$ (excluding the Ly$\alpha$ PSF, see \S3). Above the 2$\sigma$ surface brightness limit of SB$_{\rm{Ly\alpha}}= 4.8\times10^{-18}$ erg s$^{-1}$ cm$^{-2}$ arcsec$^{-2}$, we measure this nebula to have an end-to-end spatial extent of $\sim$ 442 kpc, comparable to the largest known Ly$\alpha$ nebula \cite[e.g.,][]{cantalupo14}. 

MAMMOTH-1 is associated with a relatively faint source in the broadband (source B, Figure 1). This source has an extended \heii\ and \civ\ emission in our LBT/MODS spectra (Figure 5). No radio sources are detected from the FIRST radio catalog (Becker et al. 1995) within 30$''$ from the center of MAMMOTH-1 (\S3). Both \civ\ and \heii\ have a spatial extent of $\gtrsim 30$ kpc. The Ly$\alpha$, \heii\ and \civ\ emission all contain two major components, with velocity offsets of $\approx700$ km s$^{-1}$ (\S4.2). 
The large spatial extent of the Ly$\alpha$, extended \heii\ and \civ\ emission, and double-peaked line profiles make MAMMOTH-1 to be unique compared to all the ELANe discovered up to date.

We discussed several explanations for MAMMOTH-1. We consider different scenarios including the photoionization (\S4.2.1), resonant scattering (\S4.2.2),  shocks due to gas flows (\S4.2.3), and  cooling radiation (\S4.2.4). 
We ruled out resonant scattering and cooling radiation as unlikely.  Our current data support  photoionization (Figure~8) or/and shocks due to the galactic outflow as the source of the extended Ly$\alpha$ emission. The outflow model could naturally generate the double-peaked structures of the \heii\ and \civ\ emission.   The future Integral Field Spectroscopy can examine if this ELAN is powered by a group of galaxies, and also can help us to better understand the nature of MAMMOTH-1.

\begin{table*} [!h]
\caption{Properties of Ly$\alpha$ nebula MAMMOTH-1} 
\label{table:PopIII_SFR}	
\centering 
\begin{tabular}{|c | c|  c|c| } 
\hline\hline 
Center$^{\bf{a}}$  &  Aperture &  $L_{\rm{total}}$ & $L_{\rm{\rm{nebula}}}$  \\ 
              &     & (10$^{43}$ erg s$^{-1})$       &  (10$^{43}$ erg s$^{-1})$                 \\  
\hline
$\alpha$ = 14:41:24.475, $\delta=$+40:03:09.45 & Entire nebula$^b$ & $52.8\pm2.0$ & $49.0\pm1.0$  \\
\hline
\hline
\end{tabular}
      \small\\
 {\bf a:} We apply source B's position as the center of MAMMOTH-1 (see Figure~1 and Figure~2). \\
     {\bf b:} 
   We include all the continuous area with surface brightness (SB) $>4.8\times10^{-18}$  erg s$^{-1}$ cm$^{-2}$ arcsec$^{-2}$).  
\end{table*}

\begin{table*} [!h]
\caption{Surface brightness of emission lines in MAMMOTH-1 Nebula (blue apertures in Figure~5)} 
\label{table:PopIII_SFR}	
\centering 
\begin{tabular}{|c | c| c |c | c|} 
\hline\hline 
 Aperture &  $SB_{\rm{total}}$ & $SB_{\rm{CIV}}$ & $SB_{\rm{HeII}}$ & $SB_{\rm{CIII]}}$  \\ 
   & (erg s$^{-1}$ cm$^{-2}$ arcsec$^{-2}$) &  (erg s$^{-1}$ cm$^{-2}$ arcsec$^{-2}$) &  (erg s$^{-1}$ cm$^{-2}$ arcsec$^{-2}$) &  (erg s$^{-1}$ cm$^{-2}$ arcsec$^{-2}$)          \\  
\hline
2$"\times 1.2"$ (16.7 $\times$10 kpc$^2$)  & $29.9\pm0.1\times10^{-17}$  & $3.7\pm0.1\times10^{-17}$ & $3.3\pm0.1\times10^{-17}$ & $1.0\pm0.1\times10^{-17}$\\ 
\hline
\hline
\end{tabular}
\end{table*}

\begin{table*}[!h]
\caption{A Comparison between the Ly$\alpha$ nebula MAMMOTH-1 and other enormous Ly$\alpha$ nebulae (ELANe)} 
\label{table:PopIII_SFR}	
\centering 
\begin{tabular}{|c | c| c |c | c| c | c |} 
\hline\hline 
Name  &  $L_{\rm{total}}$ & $L_{\rm{\rm{nebula}}}$ & size$^a$ &  Ionizing sources & $L_{\rm{CIV}}/L_{\rm{Ly\alpha}}$ & $L_{\rm{HeII}}/ L_{\rm{Ly\alpha}}$\\ 
  & (10$^{43}$ erg s$^{-1})$ &  (10$^{43}$ erg s$^{-1})$ & (kpc) &       &  &          \\  
\hline
\hline
MAMMOTH-1 & $52.8\pm0.1$  & $49.0\pm1.0$  &  $\approx440$ &  faint source   & $0.12\pm0.01$ & $0.12\pm0.01$\\
                      &                       &                           &  & ($U=25.8$, $B= 23.7$, $V=24.3$, $i=24.3$) & &  \\  
\hline
Slug$^b$ &  143.0 $\pm5.0 $ & $\approx 22.0$ & $\approx 500$  & ultraluminous QSO  & $< 0.12$ (2$\sigma$) & $<0.08$ (2$\sigma$) \\
\hline
Jackpot$^c$ &      $\approx 20.0$            &       &      $\approx310$                     &  QSO quartet, ultraluminous QSO & & 
 \\
\hline
Q0042-2627$^d$ & $17.0$  & &  318 &  ultraluminous QSO & $<0.01\ (2\sigma)$  & $< 0.01\ (2\sigma)$
\\
\hline
CTS G18.01 $^d$  & $19.0$ &   & 239 & ultraluminous QSO & $<0.04\ (2\sigma)$ & $<0.03\ (2\sigma)$ \\ 
\hline
\hline
\hline
\end{tabular}
\footnotetext[1]{ We define size for a surface brightness SB$\ge4.8\times10^{-18}$  erg s$^{-1}$ cm$^{-2}$ arcsec$^{-2}$. For the other Ly$\alpha$ nebulae, the luminosities are just cited from the published papers, without defining a surface brightness threshold. }
\footnotetext[2]{The parameters of the Slug nebula are  from Cantalupo et al. (2014) and Arrigoni-Batta et al. (2015). }
\footnotetext[3]{The parameters of the Jackpot nebula are  from \citep{hennawi15}. }
\footnotetext[4]{The parameters of the two Ly$\alpha$ nebulae, Q0042-2627 and CTS G18.01, are  from the MUSE Ly$\alpha$ nebulae survey (Borisava et al. 2016). }
\end{table*}

{{\bf Acknowledgements: }  ZC acknowlendges the valuable comments from Fabrizio Arrigoni Battaia, Joe Hennawi and Arjue Dey. ZC, XF, and IM thank the support from the US NSF grant AST 11-07682. ZC and JXP acknowledge support from NSF AST-1412981. AZ acknowledges support from NSF grant AST-0908280 and NASA grant ADP-NNX10AD47G. SC gratefully acknowledges support from Swiss National Science Foundation grant PP00P2\_163824. NK acknowledges supports from the JSPS grant 15H03645. Based on observations at Kitt Peak National Observatory, National Optical Astronomy Observatory (NOAO Prop. ID: 2013A-0434; PI: Z. Cai; NOAO Prop. ID: 2014A-0395; PI: Z. Cai), which is operated by the Association of Universities for Research in Astronomy (AURA) under cooperative agreement with the National Science Foundation. The authors are honored to be permitted to conduct astronomical research on Iolkam Du'ag (Kitt Peak), a mountain with particular significance to the Tohono O'odham. The LBT is an international collaboration among institutions in the United States, Italy and Germany. The LBT Corporation partners are: The University of Arizona on behalf of the Arizona university system; Istituto Nazionale di Astrofisica, Italy;  LBT Beteiligungsgesellschaft, Germany, representing the Max Planck Society, the Astrophysical Institute Potsdam, and Heidelberg University; The Ohio State University; The Research Corporation, on behalf of The University of Notre Dame, University of Minnesota and University of Virginia. }

\end{document}